\documentclass[useAMS,usenatbib]{mn2e}
\usepackage{graphicx}

\title[{\it{Strange Star Magnetospheric Activity}}]{The magnetospheric activity of bare strange quark stars}
\author[J. W Yu and R. X. Xu]{J. W. Yu and R. X. Xu\\
School of Physics and State Key Laboratory of Nuclear Physics and Technology,  Peking University, Beijing 100871, China\\
email:  J.W.Yu@pku.edu.cn,  r.x.xu@pku.edu.cn}

\begin{document}

\date{Accepted 2011 January 23. Received 2011 January 16; in original form 2010 December 05}

\pagerange{\pageref{firstpage}--\pageref{lastpage}} \pubyear{2011}

\maketitle

\label{firstpage}

\begin{abstract}

In Ruderman \& Sutherland (RS75) model, the normal neutron stars as
pulsars bear a severe problem, namely the binding energy problem
that both ions (e.g., ${}_{26}^{56}$Fe) and electrons on normal
neutron star surface can be pulled out freely by the unipolar
generator induced electric field so that sparking on polar cap can
hardly occur. This problem could be solved within the Partially
Screened Gap (PSG) model in the regime of neutron stars. However, in
this paper we extensively study this problem in a bare strange quark
star (BSS) model. We find that the huge potential barrier built by
the electric field in the vacuum gap above polar cap could usually
prevent electrons from streaming into the magnetosphere unless the
electric potential of a pulsar is sufficiently lower than that at
infinite interstellar medium. Other processes, such as the diffusion
and thermionic emission of electrons have also been included here.
Our conclusions are as follows: both positive and negative particles
on a BSS's surface would be bound strongly enough to form a vacuum
gap above its polar cap as long as the BSS is not charged (or not
highly negative charged), and multi-accelerators could occur in a
BSS's magnetosphere. Our results would be helpful to distinguish
normal neutron stars and bare quark stars through pulsar's
magnetospheric activities.

\end{abstract}
\begin{keywords}
 magnetospheric activities---pulsars: general
\end{keywords}

\section{Introduction}

Although pulsar-like stars have many different manifestations, they
are populated mainly by rotation-powered radio pulsars.
A lot of information about pulsar radiative process is inferred from the
integrated and individual pulses, the sub-pulses, and even the
micro-structures of radio pulses.
Among the magnetospheric emission models, the user-friendly nature
of Ruderman \& Sutherland (1975; hereafter RS75) model is a virtue
not shared by others~\citep{Shukre92}.

In RS75 and its modified versions~\citep[e.g.,][]{QL98}, a vacuum gap
exists above polar cap of a pulsar, in which charged particles
(electrons and positrons) are accelerated because of ${\bf E \cdot
B} \neq 0 $.
These accelerated charged particles, moving along the curved
magnetic field lines, radiate curvature or
inverse-Compton-scattering-induced high energy photons which are
converted to $e^\pm$ while propagating in strong magnetic field.
A follow-up breakdown of the vacuum gap produces secondary
electron-positron pairs plasma that radiate coherent radio emission.
These models with gap-sparking provide a good framework to analyze
observational phenomena, especially the
drifting~\citep{DC68,DR99,VJ99} and bi-drifting~\citep{QLZXW2004}
sub-pulses.

However, the RS75-like vacuum gap models work only in strict
conditions: strong magnetic field and low temperature on surface of
pulsars~\citep[e.g.,][]{GH08,ML07}.
% with ${\bf \Omega \cdot B}<0$~\citep[e.g.,][]{GH08,ML07},
%where $\bf \Omega$ is the angular frequency and $\bf B$ is the
%magnetic field.
%
The necessary binding energy of positive ions (e.g.,
${}_{26}^{56}$Fe) for RS75 model to work should be higher than $\sim
10$ keV, while calculations showed that the cohesive energy of
${}_{26}^{56}$Fe at the neutron star surface is $<1$
keV~\citep{FLRSHM77,L01}.
This binding energy problem could be solved within a partially
screened inner gap model~\citep{GM03,GM06,MG09} for normal neutron stars.
%with ${\bf \Omega \cdot B}<0$.
%
Alternatively, it is noted that the binding energy could be
sufficiently high if pulsars
are bare strange quark stars~\citep{XQ98,XQZ99,XZQ01} although
strange stars were previously supposed to exist with
crusts~\citep{AFO86}.
Certainly, it is very meaningful in the elementary strong
interaction between quarks and the phases of cold quark matter that
the binding energy problem could be solved by bare quark stars as
pulsars~\citep{X09,X10}.

Though the ideas of solving the binding energy problem in BSS model
were presented and discussed in some literatures, there is no
comprehensive study with quantitative calculations up to now.
In this paper, we are going to investigate the BSS model in
quantitative details and show the physical picture of binding of
particles on BSS's surface. Our research results are that
multi-accelerators could occur above the polar cap for (and only
for) the curvature-radiation-induced (CR-induced) sparking normal
pulsars (NPs), but for other cases, such as resonant
inverse-Compton-scattering-induced (ICS-induced) sparking NPs and
both CR-induced and ICS-induced millisecond pulsars (MSPs),
particles on surface of BSSs are bound strongly enough to form
vacuum gap and RS75-like models work well if pulsars are BSSs.

\section{The accelerators above polar caps of bare strange quark stars}

On a BSS's surface, there are positively ($u$-quarks) and negatively
($d$- and $s$-quarks and electrons) charged particles. Quarks are
confined by strong color interaction, whose binding energy could be
considered as infinity when compared with the electromagnetic
interaction, while electrons are bound by electromagnetic
interaction. Therefore, in this paper we focus on the binding of
electrons.

Let's discuss briefly the binding of electrons in the BSS model at
first. On one hand, assuming the electric potential at the top of
RS75 vacuum gap is the same as that of the interstellar medium, one
could then have a potential barrier for electrons by integrating the
gap electric field from top to bottom in the vacuum gap. This
potential barrier could then prevent electrons streaming into
magnetosphere. On the other hand, electrons above the stellar
surface of BSS are described in the Thomas-Fermi model, in which the
total energy of eletrons on Fermi surface would be a constant,
$\phi_0$. In previous work (e.g. Alcock et al. 1986), this constant
is chosen to be zero, $\phi_0=0$, because they didn't consider the
effect of spinning BSS with strong magnetic fields. Due to the
unipolar generator effect, potential drop between different magnetic
field lines is set up from pole to equatorial plane. This potential
drop could result in different $\phi_0$, at different polar angle,
$\theta$, and the total energy of electrons would then be obtained
by choosing certain zero potential magnetic field line (i.e., at
$\theta_{\rm B}$ or $\theta_{\rm C}$ in Fig.~\ref{antipulsar}).
Finally, comparing the total energy of electrons with the height of
the potential barrier in vacuum gap, we can see whether eletrons can
stream into magnetosphere freely or not.
\subsection{The energy of electrons on Fermi surface}
The distribution of electrons in BSSs is described in the
Thomas-Fermi model~\citep{AFO86}. In this model, equilibrium of
electrons in an external electric field assures that the total
energy of each electron on Fermi surface is a constant, $\phi_0$.
For the case of extremely relativistic degenerate electron gas, it gives~\citep{AFO86}
\begin{equation}\label{FD}
 \epsilon(\vec{r})=cp_{\rm F}(\vec{r})-e\varphi(\vec{r})=\phi_{0},
\end{equation}
where $\epsilon(\vec{r})$ is the total energy, $cp_{\rm F}(\vec{r})$
is the Fermi energy, $-e\varphi(\vec{r})$ is the electrostatic
potential energy of electrons and $\phi_{0}$ is a constant,
describing the potential energy of electrons in the Thomas-Fermi
model at infinity.

On the other hand, the potential distribution of electrons on the
star's surface due to the electric field induced by the rotating,
uniformly magnetized star, for the sake of simplicity, could be
assumed and estimated as (Xu et al. 2006, Eq. 2 there)
\begin{equation}\label{UGP}
 V_{\rm i}(\theta) \simeq 3 \times 10^{16} B_{12} R_{6}^{2} P^{-1} \sin^{2} \theta~({\rm V}) +V_{0},
\end{equation}
where $B_{12}=B/(10^{12} \; {\rm G})$, and $R_{6}=R/(10^6 \;{\rm
cm})$ is the radius of a pulsar, $P=2\pi/\Omega$ is the pulsar
period, $\theta$ is the polar angle and $V_{0}$ is another constant.
In view of the distribution of electron above the surface of BSS
extends only thousands of femtometers, the macroscopic potential
drop between different magnetic field lines could be thought to be
at infinity in the Thomas-Fermi model. And the potential energy
related to Eq.~\ref{UGP}, $eV_{\rm i}$, could be regarded as the
constant, $\phi_{0}$, in Eq.~\ref{FD}. By choosing the certain zero
potential magnetic field line, we could obtain the total energy of
electrons, namely $eV_{\rm i}$. Two scenarios could be possible
here. The first scenario is that we choose the critical field lines
whose feet are at the same electric potential as the interstellar
medium~\citep{GJ69} as the zero potential. We may also suggest a
second choice that the zero potential should be at those magnetic
field lines which separate annular and core regions determined by
$S_{\rm AG}=$$S_{\rm CG}$, where $S_{\rm AG}$ and $S_{\rm CG}$, are
the stellar surface areas of annular region and core region,
respectively. The second scenario is based on the idea that if
particles with opposite charge stream into the magnetosphere with
$\rho_{\rm GJ}$ in both regions, areas of this two regions should
approximately be equal in order to keep the star not charging. The
feet of the critical field lines and the magnetic field lines
determined by $S_{\rm AG}=$$S_{\rm CG}$ are designated as C and B,
respectively (Fig.~\ref{antipulsar}). For the above two scenarios,
the total energy, $\phi_{\rm i}=eV_{\rm i}$, of electrons on the
Fermi surface are given by
\begin{equation}\label{EEA}
 \phi_{\rm i,C}(\theta) \simeq -3 \times 10^{10} B_{12} R_{6}^{2} P^{-1}(\sin^{2} \theta - \sin^{2} \theta_{\rm C} ) ~\;{\rm MeV},
\end{equation}
and
\begin{equation}\label{EEB}
 \phi_{\rm i,B}(\theta) \simeq -3 \times 10^{10} B_{12} R_{6}^{2} P^{-1}(\sin^{2} \theta - \sin^{2} \theta_{\rm B} ) ~\;{\rm MeV},
\end{equation}
respectively, where $\theta_{\rm C}$ and $\theta_{\rm B}$ are polar
angles of C and B (see Fig.~\ref{antipulsar}). Equations~\ref{EEA}
and~\ref{EEB} imply that the total energy of electrons is higher at
the poles and decreases toward the equator for an \lq antipular\rq~
($\bf \Omega \cdot B > 0$), which means that electrons in different
regions above a polar cap may behave differently.

\subsection{The potential barrier of electrons in vacuum gap}
\begin{figure}
\includegraphics[scale=0.4]{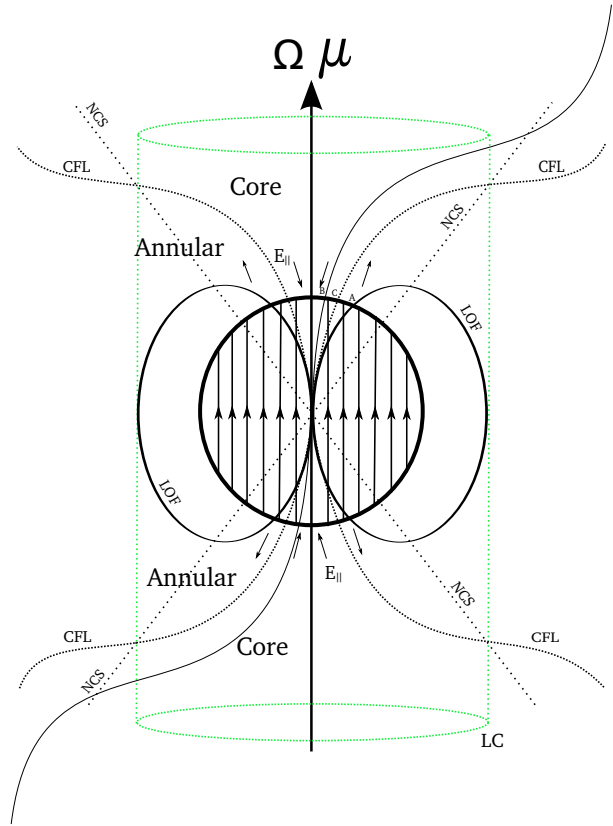}
\caption{A schematic representation of the geometry of~\lq
antipulars\rq. CFL stands for the critical field lines, NCS for null
charge surface, and LC for light cylinder. The enlarged arrows with
opposite directions in annualr region and core region represent the
directions of the electric field in vacuum gap. ``A'',``B'' and ``C''
represent the feet of different magnetic field lines (see text).}
\label{antipulsar}
\end{figure}

\begin{figure}
  \includegraphics[scale=0.3]{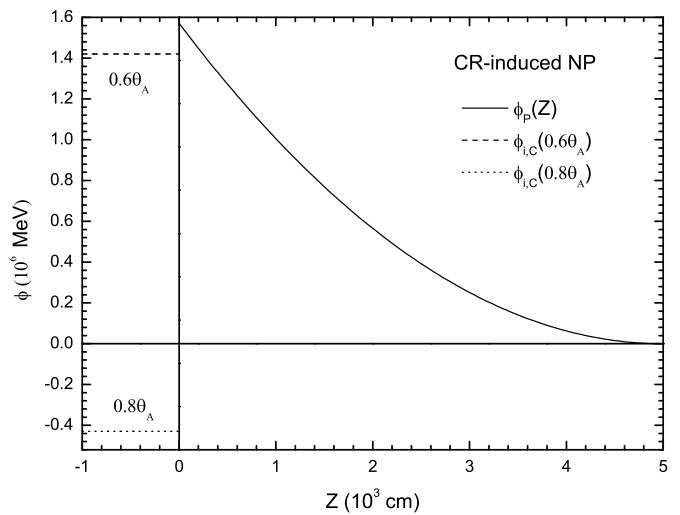}
    \caption{The potential barrier of electrons, $\phi_{\rm P}$, in vacuum gap of typical NPs ($P = 1 \:{\rm s}$, $B = 10^{12}$ G). The potential energy of electrons at stellar surface,
    namely $\phi_{\rm i}(\theta)$, is illustrated with fixed polar angles, for example, with $0.6\theta_{\rm A}$ and $0.8\theta_{\rm A}$, where $\theta_{\rm A}$ is the polar angle of
    the feet of the last open field lines (Fig.~\ref{antipulsar}).}
    \label{PB}
\end{figure}

In the following, we will consider the potential barrier of
electrons in vacuum gap. Unlike RS75, we do calculations in
situation of an \lq antipulsar\rq~whose magnetic axis is parallel to
its spin axis. A schematic representation for \lq antipulsar\rq~is
shown in Fig.~\ref{antipulsar}. Assuming the electric potential at
the top of RS75 vacuum gap is the same as that of the interstellar
medium, we could get a potential barrier for electrons by
integrating the gap electric field from top to bottom in the vacuum
gap. This potential barrier, in one-dimensional approximation, is
(RS75)
\begin{equation}\label{EEC}
 \phi_{\rm p}(Z)=2\pi \times 10^{4} P^{-1} B_{12} (h_{3}-Z_{3})^{2} ~\;{\rm MeV},
\end{equation}
where $h_{3}=h/(10^3 \:{\rm cm})$ is the height of vacuum gap,
$Z_{3}=Z/(10^3 \:{\rm cm})$ is the space coordinate measuring height
above the quark surface. This potential barrier may prevent
electrons injecting into pulsar's magnetosphere. The height of this
potential barrier mainly depends on the height of vacuum gap which
is determined by cascade mechanics of sparking, i.e., the CR-induced
cascade sparking and the ICS-induced cascade sparking.
In CR-induced cascade sparking model, the gap height is (RS75)
\begin{equation}\label{HCR}
 h_{\rm CR}=5 \times 10^{3} \rho_{6}^{2/7} B_{12}^{-4/7} P^{3/7} ~\; {\rm cm},
\end{equation}
and in ICS-induced cascade sparking model, it is~\citep{ZHM00}
\begin{equation}\label{HICS}
h_{\rm ICS}=2.79 \times 10^{4} \rho_{6}^{4/7} B_{12}^{-11/7} P^{1/7} ~\; {\rm cm}.
\end{equation}
In previous work of Gil et al. (2006), the heights of the vacuum gap
of both CR-induced and ICS-induced sparking mechanism (Gil et al.
2006, Eqs. 21 and 22 there) are different from what we used in this
work. In the PSG model, there was a partial flow of iron ions from
the positively charged polar cap which coexist with the production
of outflowing electron-positron plasmas. Such a charge-depleted
acceleration region is also highly sensitive to both the critical
ion temperature and the actual surface temperature of the polar
cap~\citep{GM03}. Differently, in our model, there is no flow of
positively charged particles, namely quarks and also it is
insensitive to the actual surface temperature. This means that there
is no partial screened effect above polar cap of bare strange quark
stars, namely the pure vacuum gap exists on polar cap of bare
strange quark stars. That's the reason why we use Eqs.~\ref{HCR}
and~\ref{HICS} in our calculation. Whether this choice of height of
vacuum gap could result in different driftrate of subpulses or not
is a complicated problem. We will discuss this problem very briefly
in \S 3. The potential barrier of electrons in the gap for
CR-induced cascade sparking model of typical normal pulsars (NPs) is
plotted in Fig.~\ref{PB}, in which the total energy of electrons at
the stellar surface, namely $\phi_{\rm i}$, is illustrated at
different polar angles. The situation of CR-induced cascade sparking
of typical millisecond pulsars (MSPs) is similar to that of NPs but
with greater height of potential barrier.

\begin{figure}
 \includegraphics[scale=0.26]{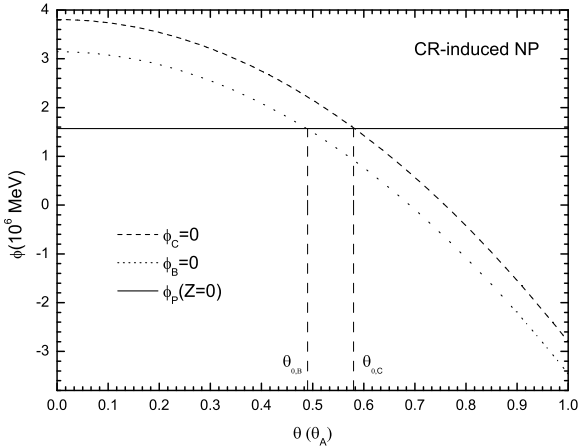}
 \caption{Comparison between the total energy of electron on stellar surface with the height of the potential barrier of typical NPs with the choice of $\phi_{\rm i}(\theta_{\rm C})=0$ and
$\phi_{\rm i}(\theta_{\rm B})=0$, respectively. The solid horizontal
line is the height of the potential barrier of electrons, namely
$\phi_{\rm P}(Z=0)$.} \label{PEC}
\end{figure}

Comparing the potential barrier with total energy of electrons, we
will explain behavior of electrons above polar cap. Namely, only
electrons with energy greater than the potential barrier can escape
into pulsar's magnetosphere. It is known that energy of electrons is
a function of polar angle (Eqs.~\ref{EEA} and~\ref{EEB}). As a
result, there may be a critical polar angle, $\theta_{0}$, at which
the energy of electrons equals the height of this potential barrier.
Comparison between the total energy of electrons and the height of
potential barrier on stellar surface for typical NPs of CR-induced
sparking is shown in Fig.~\ref{PEC} ($\theta_{0}$ does not exist for
both ICS-induced sparking of NPs and MSPs, see Table.~\ref{PA}). The
results are as follows: free flow status stays in the region of [0,
$\theta_{0}$] and vacuum gap in [$\theta_{0}$, $\theta_{\rm A}$] for
\lq antipulsars\rq,~where $\theta_{\rm A}$ is polar angle of the
feet of the last open field lines (Fig.~\ref{antipulsar}). We give
the results of $\theta_{0}$ in Table.~\ref{PA} for both pulsar and
\lq antipulsar\rq,~and find that for the special case of CR-induced
sparking NPs, free flow and vacuum gap could coexist above polar cap
which differs from the previous scenario. The general case is that
only vacuum gap exists.
\begin{table*}
 \centering
 \begin{minipage}{140mm}
 \caption{The polar angles of $\theta_{\rm B}$, $\theta_{\rm C}$ and $\theta_{0}$ for both CR-induced and ICS-induced sparking of typical NPs and MSPs within both the choice of zero potentials.\label{PA}}
 \begin{tabular}{@{}lllllllll@{}}
 \hline
    &  & &  & \multicolumn{2}{c}{$\theta_{\rm 0,B}$~($\theta_{\rm A}$)} & \multicolumn{2}{c}{$\theta_{\rm 0,C}$~($\theta_{\rm A}$)}  &  \\
  \cline{5-8}\\
   &  $\theta_{\rm A}$~(rad)  &  $\theta_{\rm B}$~($\theta_{\rm A}$) &  $\theta_{\rm C}$~($\theta_{\rm A}$)  & CR & ICS & CR & ICS & \\
 \hline
  & & &  &0.49  &...$^1$  &0.58  &... &$\bf \Omega \cdot B > 0$  \\
  NPs  & 0.0145 & 0.69 &0.76 &0.84 &2.76$^2$  &0.90 &2.83$^2$& $\bf \Omega \cdot B < 0$ \\
  &  &  &   &...   &...  &...    &... &$\bf \Omega \cdot B >0 $ \\
   MSPs & 0.145 &0.69  &0.76 &1.49$^2$ &... &1.52$^2$ &... &$\bf \Omega \cdot B <
   0$\\
 \hline
 \end{tabular}\\
$^1${$\theta_{0}$~does not exit, which means that the whole polar
cap region is vacuum gap.}\\
 $^2${$\theta_{0}$~$>$~$\theta_{\rm A}$,~which means that the whole polar cap region is vacuum gap.}
 \end{minipage}
 \end{table*}
 \subsection{The effects of thermionic emission and diffusion of electrons}
It follows from the previous argument that electrons inside BSSs
usually cannot stream into magnetospheres. Does any other process
which may affect the existence of vacuum gap above polar cap? In
vacuum gap, except pulling electrons from the interior of BSSs, two
other processes which may also prevent vacuum gap from being formed
are required to be investigated. One is the thermionic emission of
electrons and another is the diffusion of electrons from the outer
edge to the inner region of polar cap. For the first one, if the
current density due to thermionic emission of electrons is much
smaller than that of Goldreich-Julian charge density, the vacuum gap
could be maintained as well. This current density is determined by
the Richard-Dushman equation~\citep{UM95}
\begin{equation}
 J_{\rm th}= 1.2 \times 10^{14} T_{6}^{2} \exp(-1.161 \times
10^{4}T_{6}^{-1}\phi_{\rm MeV})\; {\rm A \:cm^{-2}},
\end{equation}
where $m_{\rm e}$ is the mass of electron, $k_{\rm B}$ is the
Boltzmann constant, $T_{6} = T/(10^6 \:{\rm K})$ is the temperature
and $\phi_{\rm MeV}=\phi/{\rm MeV}$ is the work function of
electrons. In the vacuum gap of BSSs, the work function of
thermionic electrons is the order of the difference between the
height of the potential barrier and the total energy of electron at
the surface of BSSs. The order of the difference is about $10^6$
MeV. At the same time, the surface temperature of polar caps of BSSs
is order of $10^6$ K. Thus, the thermionic emission current density
is $\sim 0$, which means that the thermionic emission of electrons
cannot affect the existence of the vacuum gap.

\begin{figure}
\includegraphics[scale=0.3]{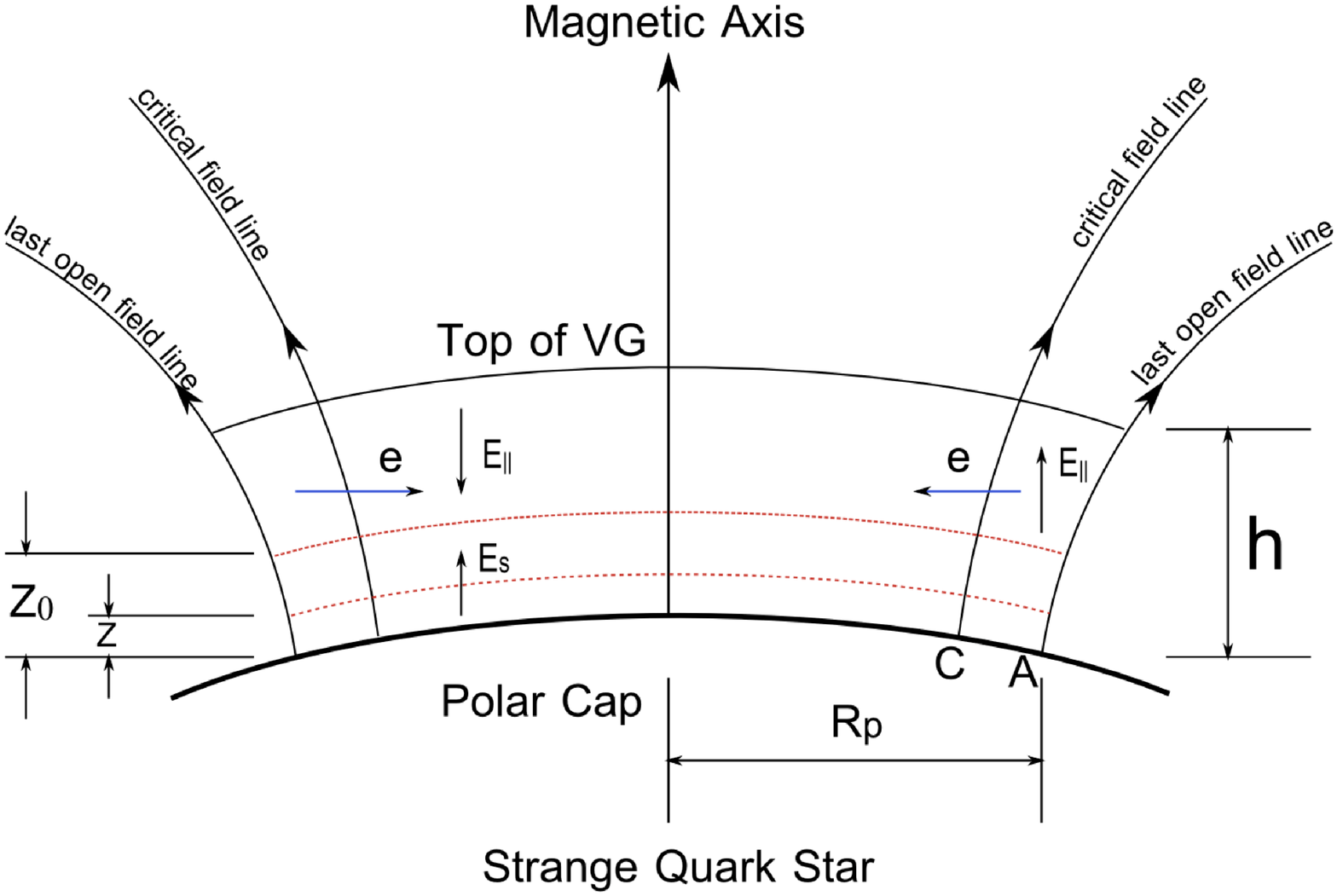}
\caption{A representative illustration of the diffusion of electrons above the
polar cap of bare strange quark star.}\label{DF}
\end{figure}

The second process is the diffusion of electrons whose distribution above BSSs surface is~\citep{XZQ01}
\begin{equation}\label{EN}
 n_{\rm e}(Z)= \frac{1.187 \times 10^{32} \phi_{{\rm q},{\rm MeV}}^{3}}
 {(0.06 \phi_{{\rm q},{\rm MeV}} Z_{11} +4)^{3}}   ~\;{\rm cm^{-3}}. %
\end{equation}
Eq.~\ref{EN} implies that the number density of electrons (so does
the kinetic energy density, $\epsilon_{\rm k}$) decreases rapidly
with increasing of the distance from quark matter surface at which
$\epsilon_{\rm k}$ $\gg$~$\epsilon_{\rm B}$, where $\epsilon_{\rm
B}$ is the magnetic field energy density.
As a result, there is a balance surface where the kinetic energy
density equals the magnetic energy density. Below this balance
surface, electrons can cross magnetic field lines freely and above
the balance surface, this motion is prevented. The physical picture
of the diffusion of electrons is illustrated in Fig.~\ref{DF}.
Making use of
$\epsilon_{\rm k} = \epsilon_{\rm B}$, where $\epsilon_{\rm k} =
n_{\rm e} \epsilon_{\rm F}$ ($\epsilon_{\rm F}$ is the Fermi energy
of degenerate electrons) and $\epsilon_{\rm B} = B^{2}/{8 \pi}$, we
can obtain the height of the balance surface.
For NPs, it is $Z_{11}$$\simeq 160$ and for MSPs, it is
$Z_{11}$$\simeq 1.7 \times 10^4$, where $Z_{11}=Z/(10^{-11} \:{\rm
cm})$. Keep in mind that there is a directed outward surface
electric field above the quark matter surface. This surface electric
field is much stronger than the gap electric field but decreases
rapidly with also increasing of the distance. Which means that this
surface electric field becomes smaller than the gap electric field
above some certain distance, $Z_{0,11}$. For NPs, it is
$Z_{0,11}$$\simeq 7000$, and for MSPs, it is $Z_{0,11}$$\simeq$$6.3
\times 10^{6}$ (see Fig.~\ref{DF}). Both have $Z_{11}$$\ll$~$Z_{0,11}$ for NPs and MSPs.
The diffusion of electrons beneath $Z_{0,11}$ is still confined by
the surface electric field meaning that only the diffusion of
electrons above the surface with height of $Z_{0,11}$ needs to be
considered.
The diffusion coefficient, $D_{\rm c}$, is given by~\citep{XZQ01}
\begin{equation}\label{DC}
 D_{\rm c} \simeq \frac{\rho^{2}}{\tau_{\rm F}} = \frac{\pi n_{\rm e} ce^{2}}{B^{2}}
  = 2.17 \times 10^{-3} B_{12}^{-2} n_{\rm {e},29}  ~\;{\rm cm^{2} \,s^{-1}},
\end{equation}
where $\rho = \gamma \rho_{\rm L}$ ($\rho_{\rm L}=m_{\rm e}vc/(eB)$ is the Larmor radius) is the
cyclotron radius of relativistic electrons, and
$\tau_{\rm F} \simeq \gamma m_{\rm e}^{2} v^{3}/(\pi e^{4} n_{\rm e})$ is the mean free flight time of electrons.
The gradient of electrons along with the diffusion direction is approximately
\begin{equation}\label{DG}
\frac{{\rm d} n_{\rm e}}{{\rm d} x} \simeq  \frac{n_{\rm e}}{\rho} = 1.4 \times 10^{38} n_{{\rm e},29}^{2/3} ~\;{\rm cm^{-4}}.
\end{equation}
Then, the diffusion rate is
\begin{equation}\label{DR}
 I_{\rm df} = 2.77 \times 10^{29} B_{12}^{-1} P^{-1/2} R_{6}^{3/2}
 \int_{Z_{0,11}}^{\infty} n_{{\rm e},29}^{5/3} \;{\rm dZ_{11}} ~\;{\rm s^{-1}}, %
\end{equation}
where $n_{{\rm e},29}=n_{\rm e}/(10^{29}\;{\rm cm^{-3}})$. For both
the NPs and MSPs with different $\phi_{\rm q}$, we give the results
of the diffusion rate $I_{\rm df}$ and $I_{\rm GJ}$ in
Table~\ref{DRV} in which the flow with the Goldreich-Julian flux is
$I_{\rm GJ}=$~$\pi$$r_{\rm p}^{2}$c$n_{\rm GJ} $~$\simeq$$1.4
\times$$10^{30}$$P^{-2}$$R_{6}^{3}$ $B_{12}$~$\rm s^{-1}$. we can
know that both have~$I_{\rm df}$~$\ll$~$I_{\rm GJ}$~for NPs and MSPs
from Table \ref{DRV}. This means that the diffusion of electrons is
also negligible which guarantees the existence of vacuum gap.
\begin{table*}
\centering
 \begin{minipage}{140mm}
\caption{ The typical value of the diffusion rate for NPs and MSPs
with different choice of $\phi_{\rm q}$. \label{DRV}}
\begin{tabular}{@{}ccccc@{}}
\hline
   &\multicolumn{2}{c}{NPs} &\multicolumn{2}{c}{MSPs} \\
  \cline{2-5}\\
 $\phi_{\rm q}$ (MeV) &$I_{\rm df}$ ($10^{24} \; {\rm s^{-1}}$)  & $I_{\rm GJ}$ ($10^{24}\; {\rm s^{-1}}$)  & $I_{\rm df}$ ($10^{17} \; {\rm s^{-1}}$) &$I_{\rm GJ}$ ($10^{17}\; {\rm s^{-1}}$) \\
\hline
  1& $\sim$4.75 & &$\sim$7.52& \\
  10&$\sim$4.91& $\sim$$1.4 \times 10^3$& $\sim$7.52&  $\sim$$1.4 \times 10^{13}$\\
  20&$\sim$4.93 & & $\sim$7.53& \\
\hline
 \end{tabular}
\end{minipage}
\end{table*}

\section{ Conclusions and Discussions}
In RS75 model, the binding energy problem is one of the most serious
problems in the normal neutron star model of pulsars. Arons and
Scharlemann (1979) developed an alternative model, the space-charge
limited flow (SCLF) model, in which the particles, both iron ions
and electrons can be pulled out freely, and form a steady
flow~\citep{AS79}. In this SCLF model, the drifting sub-pulse
phenomenon which has been commonly observed in pulsars can rarely be
reproduced. The prerequisite for understanding this phenomenon could
be the existence of a vacuum gap.

In a very special case, through our calculations, we find that there
is a new physical scenario for CR-induced sparking of normal pulsars
(NPs) that free flow and vacuum gap may coexist above the polar cap.
But in other cases, such as ICS-induced sparking of NPs and
millisecond pulsars (MSPs), only vacuum gap exists. In general, if a
pulsar is not highly negatively charged~\citep{XCQ06}, vacuum gap
survives at polar cap as well. One limitation is that our
calculation is based on one-dimensional approximation and it might
fail in some cases of MSPs. As far as we find, it is very difficult
to deal with the high-dimensional cases. The one-dimensional
approximation provides a good understanding of the geometry of polar
cap of BSSs. In conclusion, the binding energy problem could be
solved completely in the BSS model of pulsar as long as BSSs are
neutral (or not highly negative charged), and the structure of polar
cap of BSSs are very different with respect to that of NSs. Detailed
information about the geometry of BSS's polar cap is given in
Table~\ref{AG}.
\begin{table*}
\centering
 \begin{minipage}{140mm}
\caption{The accelerators above polar caps of BSSs. \label{AG}}
\begin{tabular}{@{}llllll@{}}
\hline
   &\multicolumn{2}{c}{[0,~$\theta_{\rm 0}$]$^{\dag}$} &\multicolumn{2}{c}{[$\theta_{\rm 0}$,~$\theta_{\rm A}$]} & \\
  \cline{2-5}\\
  &CR &ICS &CR &ICS & \\
\hline
   &SCLF  &VG &VG  &VG &$\bf \Omega \cdot B > 0$ \\
  NPs  &VG &VG$^{\ddag}$ &SCLF &VG$^{\ddag}$& $\bf \Omega \cdot B < 0$ \\
  &VG   &VG  &VG    &VG &$\bf \Omega \cdot B >0 $ \\
   MSPs &VG$^{\ddag}$ &VG &VG$^{\ddag}$ &VG &$\bf \Omega \cdot B < 0$\\
\hline
 \end{tabular}\\
 $^{\dag}${$\theta_{0}$ represents $\theta_{\rm 0,B}$ while choosing $\phi_{\rm B}=0$  and  $\theta_{\rm 0,C}$ while choosing  $\phi_{\rm C}=0$.}\\
$^{\ddag}${for such cases, $\theta_{0}$ $>$~$\theta_{\rm A}$, which
represents the structure of the whole polar cap region.}
\end{minipage}
\end{table*}
A more interesting region from pole to equator may locate between
that polar angle where the total energy of electron equals the
potential barrier and the polar angle of the foot of zero potential
magnetic field line (i.e., [$\theta_{0, \rm C},\theta_{\rm C}$] or
[$\theta_{0, \rm B},\theta_{\rm B}$], see Fig.~\ref{PEC}) for
CR-induced sparking NPs. After the birth of a NP, a vacuum gap
exists at this region. When sparking starts, the potential in vacuum
gap drops rapidly due to screen by electron-positron pairs and may
become lower than that at the surface, namely $V_{\rm i}(\theta)$.
As a result, the sparking converts vacuum gap to free flow at this
region until the sparking ends, i.e., at [$\theta_{0, \rm
C},\theta_{\rm C}$] or [$\theta_{0, \rm B},\theta_{\rm B}$], vacuum
gap and free flow work alternately. This argument may have profound
implications for us to distinguish neutron stars and quark stars by
pulsar's magnetospheric activities (e.g., the diversity pulse profiles).

Another issue to be discussed is about the drifting rate of
subpulses when we use the height of pure vacuum gap in this work.
The natural explanation of the drifting subpulse phenomena in vacuum
gap is due to $\bf E \times B$. Unfortunately, these theoretical
calculations gave higher drifting rate with respect to
observations~\citep[e.g.,][]{RS75, DR99, DR01, GM03, GMZ06}. Since
it has been observed~\citep{DC68}, the drifting subpulse phenomenon
remains unclear which has been widely regarded as one of the most
critical and potentially insightful aspects of pulsar
emission~\citep{DR01}. The PSG mechanism~\citep[e.g.,][]{GM03, GM06,
GMZ06} could be a way to understand lower drifting rates observed,
but some complexities still exist which make the underlying physics
of drifting subpulses keep complicated and far from knowing clearly.
(1) in principle the drifting velocity of subpulses is the ratio of
the drifting distance to the duration, while the expected velocity
predicted by $\bf E \times B$ is only for electrons in separated
emission units, namely the plasma filaments. These two velocities
would not be the same if the plasma filaments may stop after
sparking. When sparking starts, the electric field in the vacuum gap
vanishes due to screen by plasmas; while sparking ends, the electric
field appears again. Thus, the calculated drifting velocity with
$\bf E \times B$ could be higher than that of observations. (2) the
so-called aliasing effect: as one observes subpulses only once every
rotation period, we can hardly determine their actual speed. The
main obstacles in the aliasing problem are the under sampling of
subpulse motion and our inability to distinguish between subpulses
especially when the differences between subpulses formed by various
subbeams are smaller than the fluctuations in subpulses from one
single subbeam~\citep{LSRR03}. Anyway, detailed studies are very
necessary in the future works.

We assume that the potential energy related to Eq.~\ref{UGP},
$eV_{\rm i}$, to be the constant, $\phi_{0}$, in Eq.~\ref{FD}. This
assumption could be reasonable. For an uniformly magnetized,
rotating conductor sphere, the unipolar generator will induce an
electric field which is a function of polar angle, as described in
Eq.~\ref{UGP}. In the case of $\bf \Omega \cdot B>0$
(Fig.~\ref{antipulsar}), the potential energy of electron is highest
at the polar region which means that those electrons there could be
easier to escape. Alternatively, this conclusion could be
quantitatively understood as following: because of Lorentz force
inside a star, more electrons locate at the polar region so that the
Fermi energy of electron is higher there and easier to escape into
magnetosphere.

\section*{Acknowledgments}
We thank Dr. Kejia Lee and other members in the pulsar group of
Peking University for their helpful and enlightened discussions.
We also thank Prof. Janusz Gil for his helpful comments and suggestions.
Junwei Yu is grateful to Dr. Caiyan Li for her helpful assistance.
The work is supported by NSFC (10973002,10935001), the National Basic
Research Program of China (grant 2009CB824800)  and the John Templeton Foundation.

\label{lastpage}
\end{document}